\documentclass[12pt]{article}
\usepackage[dvips]{graphics}
\usepackage{cite}
\usepackage{epsfig}

\catcode`@=11
\def\@citex[#1]#2{\if@filesw\immediate\write\@auxout{\string\citation{#2}}\fi
  \def\@citea{}\@cite{\@for\@citeb:=#2\do
    {\@citea\def\@citea{,\penalty\@m}\@ifundefined
      {b@\@citeb}{{\bf ?}\@warning
       {Citation `\@citeb' on page \thepage \space undefined}}%
\hbox{\csname b@\@citeb\endcsname}}}{#1}}

\def\citer{\@ifnextchar
[{\@tempswatrue\@citexr}{\@tempswafalse\@citexr[]}}
\def\@citexr[#1]#2{\if@filesw\immediate\write\@auxout{\string\citation{#2}}\fi
  \def\@citea{}\@cite{\@for\@citeb:=#2\do
    {\@citea\def\@citea{--\penalty\@m}\@ifundefined
       {b@\@citeb}{{\bf ?}\@warning
       {Citation `\@citeb' on page \thepage \space undefined}}%
\hbox{\csname b@\@citeb\endcsname}}}{#1}}
\catcode`@=12
\newcommand{\be}{\begin{equation}}
\newcommand{\ee}{\end{equation}}
\newcommand{\bea}{\begin{eqnarray}}
\newcommand{\eea}{\end{eqnarray}}
\newcommand{\ba}{\begin{array}}
\newcommand{\ea}{\end{array}}

\renewcommand{\Large}{\large}

\begin{document}

\hfill \\ \vspace{10pt} 
\hfill {\normalsize UPR--927--T} \\

\vspace{15pt}

\centerline{\Large\bf Fundamental Parameters from Precision Tests\footnote{Talk
presented at the Symposium in Honor of Alberto Sirlin, NYU, NY, October 2000.}}

\vspace{15pt}

\centerline{\sc Jens Erler} 

\vspace{15pt}

\centerline{\it 
            Department of Physics and Astronomy, University of Pennsylvania,}
\centerline{\it Philadelphia, PA 19104-6396}
\centerline{\it E-mail: erler@langacker.hep.upenn.edu}

\begin{abstract}
I review how electroweak precision observables can be used to constrain
the Higgs boson mass and the strong coupling constant.  Implications for
physics beyond the Standard Model are also addressed.
\end{abstract}

\nopagebreak

\section{Introduction}
Unlike most speakers before me, I never collaborated with Alberto Sirlin 
directly. However, my work and the way I look at precision tests of electroweak
physics are heavily influenced by his work. In fact, it was the clarity of his 
papers (I have listed my all time favorites in Ref.~\cite{Alberto}), which 
finally encouraged me to accept the task to write an independent FORTRAN 
package, to be used for the {\sl Global Analysis of Particle Properties} 
(GAPP)~\cite{Erler:1999ug}. It is a pleasure to be part of this Symposium in 
his honor and to meet many of his collaborators, some of them for the first 
time.  

My talk should be understood as a continuation of Paul Langacker's 
contribution~\cite{Paul}, in which he reviewed the history of electroweak 
physics until today.  Using the latest set of data he presented, I want 
to discuss the determination of various parameters within and beyond 
the Standard Model (SM). I will spend some time talking about the extraction of
the strong coupling constant, $\alpha_s$, from electroweak processes, and I 
will give some details regarding the current constraints on the Higgs boson 
mass, $M_H$. Finally, I will briefly discuss oblique parameters, which are 
relevant to a specific class of new physics.

Paul showed a long list of observables (his Table~1) of current $Z$-pole 
experiments. In view on these results, one feels an obligation to stress 
the impressive agreement between the individual experiments and the SM 
predictions. In fact, there are only two $Z$ pole results which deviate by more
than $1.5\sigma$ from the SM, namely the hadronic peak cross section 
($1.7\sigma$), and the forward-backward cross section asymmetry for bottom 
quark final states, $A_{FB}(b)$ ($2.5\sigma$). 

Paul also showed a Table (his Table~2) of non $Z$ pole experiments. Unlike 
a few months ago when the largest SM deviation occurred in the effective weak 
charge of Cs, $Q_W({\rm Cs})$, a new atomic structure theory 
calculation~\cite{Derevianko:2000dt} now implies virtual agreement ($1\sigma$).
Note, that a large deviation in $Q_W$ could indicate the presence of an extra 
neutral gauge boson~\cite{Erler:2000nx}.

\begin{figure}[t]
\begin{center}
\mbox{\psfig{figure=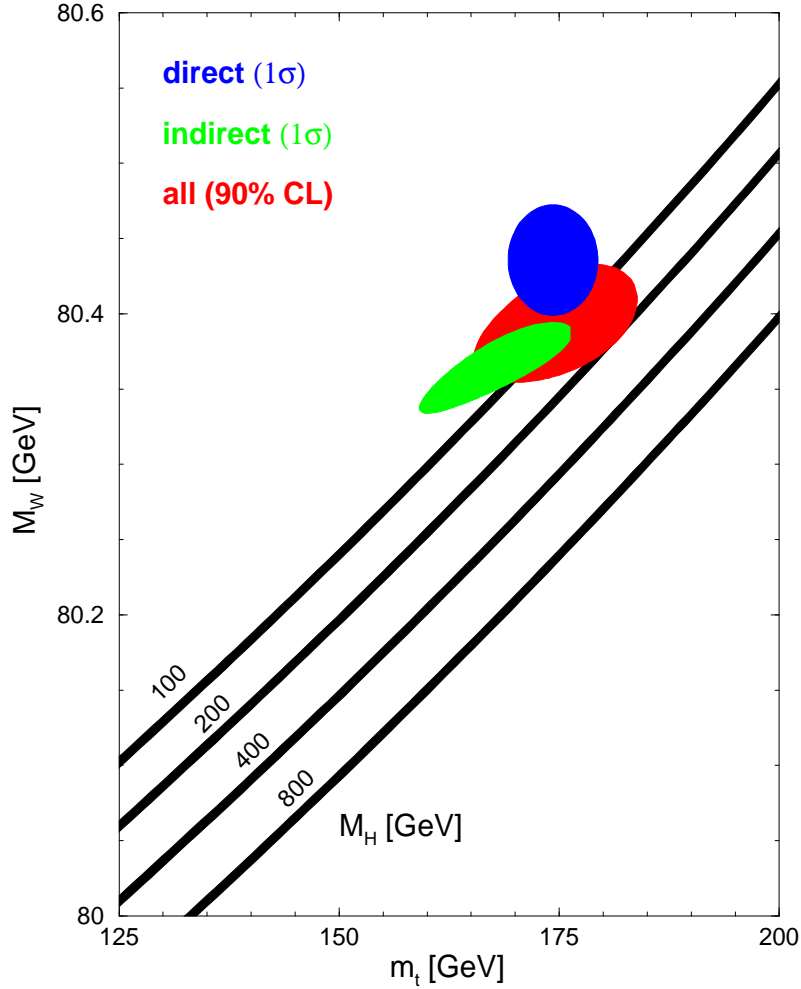,width=300pt,height=380pt}}
\end{center}
\caption{Direct and indirect determinations of $M_W$ and $m_t$. The bands are 
the SM predictions for various values of $M_H$, their widths indicating
the uncertainties from other inputs.}
\label{mwmt}
\end{figure}

One way of summarizing the current situation is in the $M_W$--$m_t$ plane.
In Fig.~\ref{mwmt} the direct measurements of the $W$ boson and top quark 
masses are compared with their indirect determinations from precision tests and
with the predictions of the SM for various values of $M_H$. One can see from 
the figure that the determinations are in perfect agreement with each other,
and also with the SM provided the Higgs mass is strictly of order 100~GeV.
The latter statement will be made more precise in the next Section. As for 
other SM parameters, for example the electromagnetic coupling constant at 
the $Z$ scale, $\Delta\alpha(M_Z)$, the weak mixing angle, $\hat{s}$, 
{\em etc.}, see Paul Langacker's talk~\cite{Paul}.

\section{$M_H$}
From the precision data I obtain the result,
\be
   M_H = 86^{+48}_{-32} \mbox{~GeV},
\ee
{\em i.e.,} a 58\% determination. The central value is 27~GeV or $0.6\sigma$ 
below the direct lower limit from LEP~2, $M_H > 113.2$~GeV 
(95\% CL)~\cite{Junk00} The 90\% {\em central\/} confidence interval is 
\be
   38 \hbox{ GeV} < M_H < 173 \hbox{ GeV}.
\ee
However, for a proper upper bound one should also take into account the direct 
search results at LEP~2, {\em i.e.\/} the exclusion of a Higgs boson with mass 
below 110~GeV or so, and the observation of several candidate events consistent
with $M_H$ around 115~GeV. Making use of Bayes' theorem~\cite{Bayes},
\be
   p(M_H|{\rm data}) = \frac{p({\rm data}|M_H)p(M_H)}{p({\rm data})},
\label{Bayes}
\ee
one can compute the entire probability distribution function of $M_H$
conditional on the data and the validity of the SM. While the numerator,
$p({\rm data})$, is easily obtained by properly normalizing 
the {\em posterior\/} probability density on the left-hand side, 
the {\em prior\/} probability, $p(M_H)$, demands some extra thought:

Depending on the case at hand, the prior can
(i) contain additional information not included in the likelihood model,
(ii) contain likelihood information obtained from previous measurements, or
(iii) be chosen {\em non-informative}.
As for the present case, I choose the {\em informative\/} prior,
$p(M_H) = Q_{\rm LEP~2}~p^{\rm non-inf}(M_H)$,
where the non-informative part of the prior is,
\be
   p^{\rm non-inf}(M_H) = M_H^{-1}.
\label{noninf}
\ee
The quantity,
\be
   Q_{\rm LEP~2} = \frac{{\cal L} ({\rm data|signal + background})}
                        {{\cal L} ({\rm data|background})},
\ee
is an $M_H$ dependent summary statistic of the Higgs searches at LEP~2.
If the signal hypothesis gives a better (worse) description of the data than 
the background only hypothesis one finds a negative (positive) contribution to 
the total $\chi^2$. Note, that this is a consistent treatment also in the case 
of a large downward fluctuation of the background or even if no events are 
observed at all. See the talk by Giuseppe Degrassi for more details and 
a somewhat different perspective~\cite{Degrassi}.

The choice~(\ref{noninf}) corresponds to a flat prior in the variable 
$\ln M_H$, and there are various ways to justify it. One rationale is that 
a flat distribution is most natural for a variable defined over all the real 
numbers. This is the case for $\ln M_H$ but not $M_H^2$. Also, it seems that 
{\em a priori\/} it is equally likely that $M_H$ lies, say, between 30 and 
40~GeV, or between 300 and 400~GeV. In any case, the sensitivity of 
the posterior to the (non-informative) prior diminishes rapidly with 
the inclusion of more data. Both, $p(M_H)$ and $p^{\rm non-inf}(M_H)$, are 
improper (not integrable) distributions, but the likelihood, 
$p({\rm data}|M_H)$, constructed from the precision measurements assures
a proper posterior.

Occasionally, the Bayesian method is criticized for the need of a prior, which 
would introduce unnecessary subjectivity into the analysis. Indeed, care and 
good judgement is needed, but the same is true for the likelihood model, which 
has to be specified in any statistical model. Moreover, it is appreciated among
Bayesian practitioners, that the explicit presence of the prior can be 
advantageous: it manifests model assumptions and allows for sensitivity checks.
From the theorem~(\ref{Bayes}) it is also clear that any other method must 
correspond, mathematically, to specific choices for the prior. Thus, Bayesian 
methods are more general and differ rather in attitude: by their strong 
emphasis on the entire posterior distribution and by their first principles 
setup. 

\begin{figure}[t]
\begin{center}
\mbox{\psfig{figure=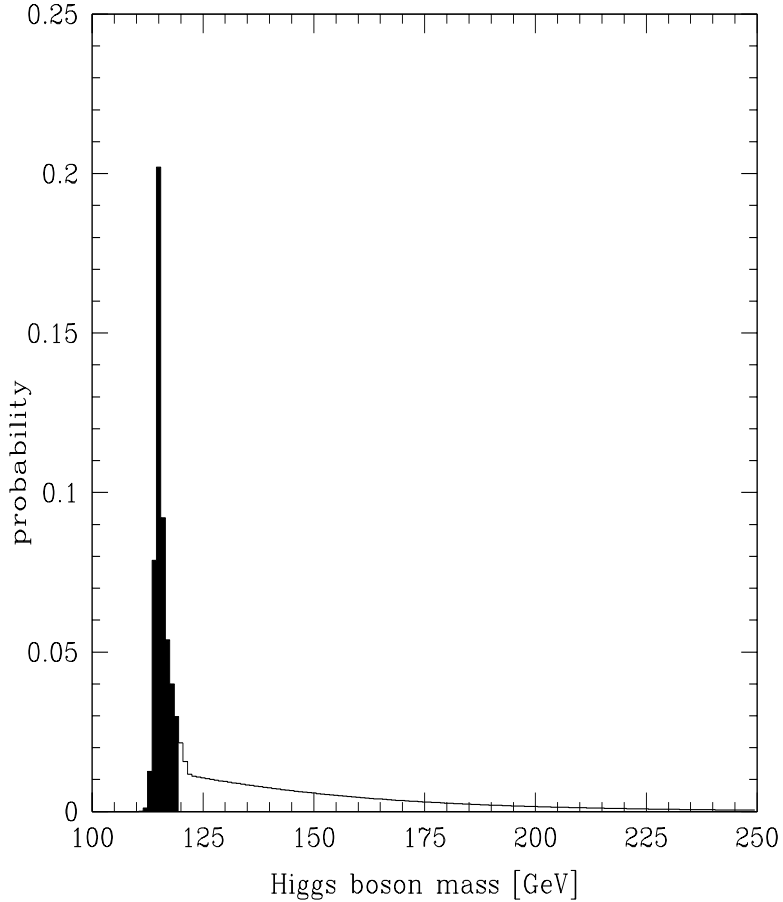,width=300pt,height=360pt}}
\end{center}
\caption[]{Probability distribution function~\cite{Erler:2000cr} for the Higgs 
boson mass. The probability is shown for bin sizes of 1~GeV. Included are all 
available direct and indirect data. The shaded and unshaded regions each mark 
50\% probability.}
\label{higgspdf}
\end{figure}

Including $Q_{\rm LEP~2}$ in this way, one obtains the 95\% CL upper limit 
$M_H \leq 201$~GeV, {\em i.e.\/} notwithstanding the observed excess events, 
the information provided by the Higgs searches at LEP~2 {\em increases\/} 
the upper limit by 28~GeV.

\begin{figure}[t]
\begin{center}
\mbox{\psfig{figure=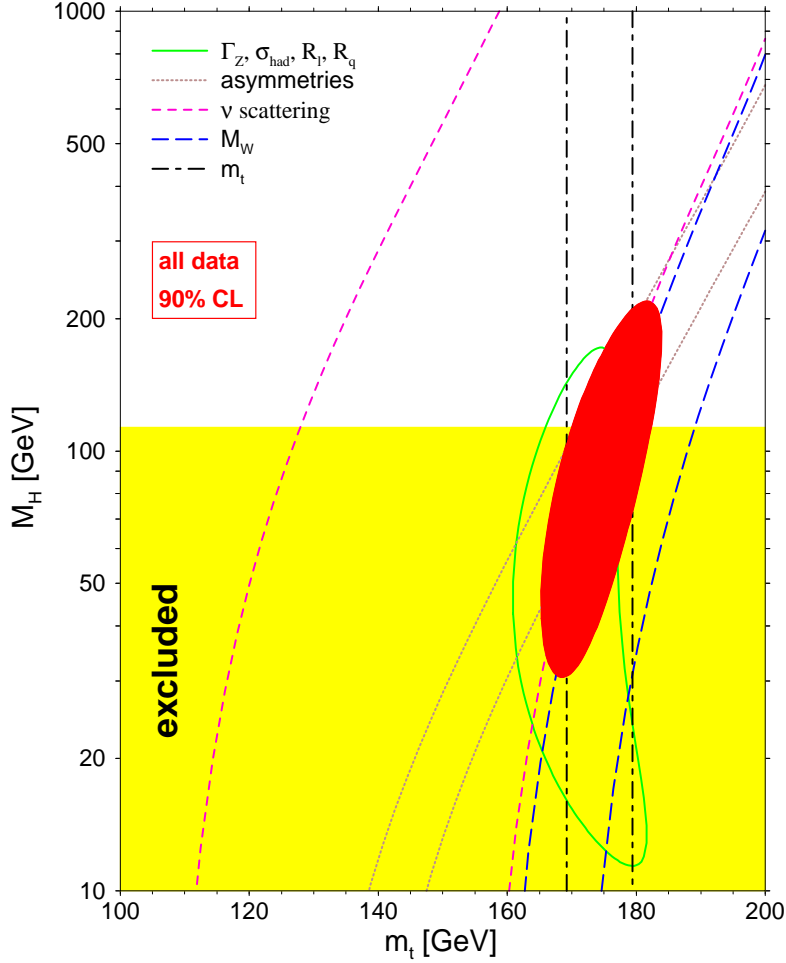,width=300pt,height=370pt}}
\end{center}
\caption{Current $1\sigma$ constraints on $M_H$ and $m_t$ from various 
precision observables. Their combination results in the red (solid) ellipse 
which contains 90\% probability. The direct limit from LEP~2 is also 
indicated.}
\label{mhmt}
\end{figure}

Given extra parameters, $\xi^i$, the distribution function of $M_H$ is defined 
as the marginal distribution, 
$p(M_H|{\rm data}) = \int p(M_H, \xi^i | {\rm data}) \prod_i p(\xi^i) d \xi^i$.
If the likelihood factorizes, $p(M_H, \xi^i) = p(M_H) p(\xi^i)$, the $\xi^i$ 
dependence can be ignored. If not, but $p(\xi^i | M_H)$ is (approximately) 
multivariate normal, then
\be
   \chi^2 (M_H,\xi^i) = \chi^2_{\rm min} (M_H) +
   {1\over 2} \frac{\partial^2 \chi^2 (M_H)} {\partial \xi_i \partial \xi_j} 
   (\xi^i - \xi^i_{\rm min} (M_H)) (\xi^j - \xi^j_{\rm min} (M_H)). 
\ee
The latter applies to our case, where $\xi^i = (m_t,\alpha_s,\alpha(M_Z))$. 
Integration yields, 
\be
   p(M_H | {\rm data}) \sim \sqrt{\det E}\, e^{- \chi^2_{\rm min} (M_H)/2},
\ee
where the $\xi^i$ error matrix, 
$E = (\frac{\partial^2 \chi^2 (M_H)} {\partial \xi_i \partial \xi_j})^{-1}$, 
introduces a correction factor with a mild $M_H$ dependence. It corresponds to 
a shift relative to the standard likelihood model given by
\be
   \Delta \chi^2 (M_H) \equiv \ln \frac{\det E (M_H)}{\det E (M_Z)}.
\ee
This effect {\em tightens} the $M_H$ upper limit by 1~GeV. I also include 
theory uncertainties from uncalculated higher orders. This increases the upper 
limit by 5~GeV and finally yields
\be
   M_H \leq 205 \mbox{ GeV} \mbox{ (95\% CL)}.
\ee

The entire probability distribution is shown in Fig.~\ref{higgspdf}. Taking
the data at face value, there is (as expected) a significant peak around 
$M_H = 115$~GeV, but more than half of the probability is for Higgs boson 
masses above the kinematic reach of LEP~2 (the median is at $M_H = 119$~GeV). 
However, if one would double the integrated luminosity and assume that 
the results would be similar to the present ones, one would find most of 
the probability concentrated around the peak. A similar statement will apply to
Run II of the Tevatron at a time when about 3 to 5 ${\rm fb}^{-1}$ of data have
been collected.

The current status of Higgs mass constraints can also be summarized graphically
in the $M_H$--$m_t$ plane. Fig~\ref{mhmt} shows the contours arising from 
the $Z$ lineshape measurements; from $Z$ pole asymmetries; from neutrino-hadron
and neutrino-electron scattering; and from $M_W$. The direct measurement of 
$m_t$ from the Tevatron, and the direct lower limit on $M_H$ from LEP~2 are 
also shown. Notice, that all groups of measurements are consistent with each 
other and the SM, provided $M_H$ is not much larger than its current lower 
limit.

\section{$\alpha_s$}
Another fundamental SM parameter is the strong coupling constant, $\alpha_s$. 
Remarkably, within the SM the cleanest determination of the QCD coupling comes 
from electroweak physics. Table~\ref{alphas} shows the constraints on 
$\alpha_s$ provided by various electroweak observables. A few comments:
\begin{itemize}
\item The quoted errors include the uncertainty from $M_H$ which is allowed
      as a free parameter.
\item The results are virtually unchanged if $M_H$ is fixed to 115~GeV instead,
      corresponding to the most likely value according to Fig.~\ref{higgspdf}.
\item The results are {\em insensitive\/} to types of new physics which couple
      preferentially to vector boson self-energies (``oblique corrections'';
      see Section~\ref{STU}). $\alpha_s = 0.1201\pm 0.0031$ allowing 
      the oblique STU parameters introduced in Ref.~\cite{Peskin:1990zt}. 
\item The extracted $\alpha_s$ is {\em very sensitive\/} to non-universal new 
      physics (to the $Zb\bar{b}$-vertex, etc.).
\item It is amusing that the electroweak fit provides a more precise value than
      the QCD average, $\alpha_s = 0.1184 \pm 0.0031$~\cite{Bethke:2000ai}.
      Authors addressing the combination of intrinsic QCD determinations of 
      $\alpha_s$ feel uncomfortable with standard statistical procedures 
      to combine information from various sources. This is mainly because 
      (i) most uncertainties are strongly dominated by theory errors which 
          oftentimes are no more than ``best guesses'';
      (ii) there may be unknown correlations and common sources of 
           uncertainties hampering straightforward averaging procedures. 
      As a reasonable but subjective countermeasure, one chooses a very 
      conservative attitude and inflates the error bars in a more or less 
      {\em ad hoc\/} manner. 
\item The global fit result in Table~\ref{alphas} {\em cannot\/} be obtained by
      averaging the individual values due to both, experimental and
      theoretical (mainly from common inputs) correlations. While these
      represent major complications, they can be addressed effectively and
      unambiguously in a global fit (unlike in the QCD world average discussed 
      before).
\end{itemize}

\begin{table}[t]
\begin{center}
\begin{tabular}{|l|r|} 
\hline
$\Gamma_Z$         & $\alpha_s =0.1182 \pm 0.0054$ \\
$\sigma_{\rm had}$ & $\alpha_s =0.1074 \pm 0.0073$ \\
$R_e$              & $\alpha_s =0.1299 \pm 0.0083$ \\
$R_\mu$            & $\alpha_s =0.1267 \pm 0.0054$ \\
$R_\tau$           & $\alpha_s =0.1160 \pm 0.0072$ \\
\hline
global fit         & $\alpha_s =0.1195 \pm 0.0028$ \\
\hline
\end{tabular}
\caption[]{$\alpha_s$ from electroweak physics. $\Gamma_Z$ is the total $Z$ 
width, $\sigma_{\rm had}$ is the hadronic cross section on the $Z$ peak, while 
the $R_\ell$ are the three ratios of hadronic to leptonic $Z$ decay widths.}
\end{center}
\label{alphas}
\end{table}

\section{$S$ and $T$}
\label{STU}
The so called oblique parameters, such as the $S$, $T$, and $U$ parameters of 
Peskin and Takeuchi~\cite{Peskin:1990zt}, are defined for types of new physics 
which 
(i) have no or negligible direct couplings to the standard fermions, and
(ii) which have an associated mass scale much larger than $M_Z$.
One example are non-standard contributions to the $\rho$ parameter, which 
measures the difference in the radiative corrections to the $W$ propagator 
relative to the $Z$ propagator. In fact, the $\rho_0$ parameter is trivially 
related to the Peskin-Takeuchi parameter, $T$, through
\be
   \rho_0 = {1\over 1 - \alpha T},
\ee
where by definition, $\rho_0 = 1$ in the SM. The $S$ parameter, measures 
essentially the difference of the contributions to the $Z$ propagator at 
$q^2 = M_Z^2$ relative to $q^2 = 0$. For example, a degenerate chiral fermion 
doublet gives a positive definite contribution to $S$ given by,
\be
   S = \sum\limits_i {[t_{3L}(i) - t_{3R}(i)]^2 \over 3\pi}
     = \frac{4 \hat{s} \hat{c}}{\hat\alpha} \frac{\Pi^{\rm new}_{ZZ}(M_Z^2)
     - \Pi^{\rm new}_{ZZ}(0)} {M_Z^2},
\ee
where $t_{3L,R}$ denote the third component of isospin for the left and 
right-handed helicities, respectively, and $\hat{s} \hat{c}/ \hat\alpha$
is a combination of gauge couplings. A global fit with $S$ allowed gives, 
\be 
   S = + 0.10^{+0.13}_{-0.23},
\ee
and simultaneously $M_H = 42^{+166}_{\,\,\, -26}$~GeV. Notice, the much larger 
error on $M_H$ compared to the SM fit.  This is because $S$ and $M_H$ are 
almost perfectly anticorrelated with a correlation of $-94\%$. Thus, in 
the presence of the $S$ parameter, the $M_H$ constraints are significantly 
weakened and we find the upper bound $M_H < 397$~GeV at the 95\% CL.

\begin{figure}[t]
\begin{center}
\mbox{\psfig{figure=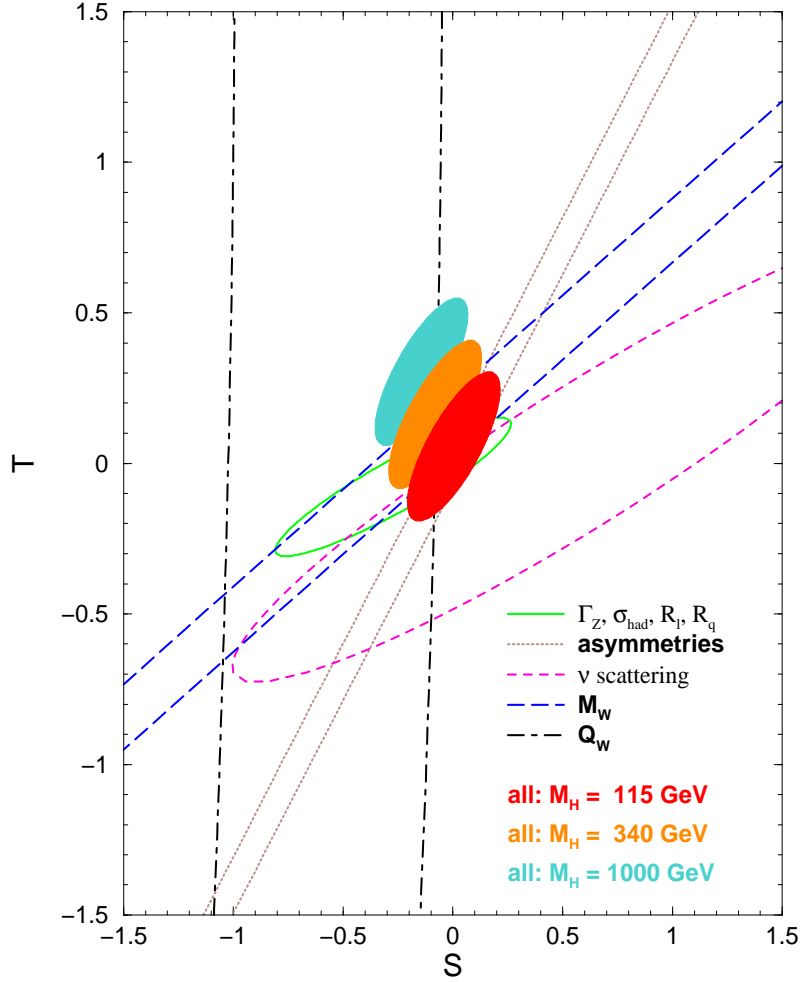,width=300pt,height=380pt}}
\end{center}
\caption[]{Current $1\sigma$ constraints on $S$ and $T$ from various 
precision observables.}
\label{ST}
\end{figure}

If $M_H$ is fixed at 115 GeV, we find $S = - 0.033 \pm 0.064$. We can use this
result to find constraints on extra degenerate fermion doublets. For example, 
a sequential or mirror family of fermions contributes 
$\Delta S = + {2\over 3 \pi}$.  From the expression
\be
   \frac{e^{-\chi^2(N_g = 3)/2}}
        {e^{-\chi^2(N_g = 3)/2} + e^{-\chi^2(N_g = 4)/2}},
\ee
we can therefore conclude that a degenerate 4th family is excluded at 
the 99.92\% CL. Or we can turn things around and compute the number of extra 
families from the $S$ parameter, $N_g = 2.84 \pm 0.30$. This is complementary 
to $N_\nu = 2.986 \pm 0.008$ arrived at from neutrino counting at LEP~1, which 
now shows a deviation of $1.7\sigma$ from the SM prediction of $N_\nu = 3$. 
In essence, this is a consequence of the measured $\sigma_{\rm had}$ which 
deviates by the same amount.

The results on the $S$ and $T$ parameters (with $U=0$) are summarized in 
Fig.~\ref{ST}, again broken down into various subsets of observables: the $Z$ 
lineshape measurements; $Z$ pole asymmetries; neutrino scattering; $M_W$; and 
$Q_W$. It is seen that the precision data is consistent with the SM prediction,
$S = T = 0$, in particular if $M_H$ is relatively light. 

One also sees that the $S$ parameter is strongly correlated with $T$. Repeating
the fit with $T$ allowed yields $N_g = 3.02 \pm 0.46$ and $T = 0.06 \pm 0.12$. 
Thus, the constraints are weaker for a {\em non-degenerate\/} 4th family.

\section*{Acknowledgement:}
It is a pleasure to thank the organizers for the invitation to a very memorable
Symposium and Paul Langacker for collaboration.

\end{document}